\documentclass{aa}  
\usepackage{graphicx}
\usepackage{txfonts}
\usepackage{lipsum}
\usepackage{subcaption}
\usepackage{lscape}
\usepackage{placeins}

\usepackage[colorlinks=true, linkcolor=blue, citecolor=blue, filecolor=blue, urlcolor=blue]{hyperref}

\newcommand{\sep}{$\Delta r$}

\newcommand{\xir}{$\xi(\Delta r)$}
\newcommand{\Xir}{$\Xi(\Delta r)$}
\newcommand{\xiv}{$\xi(\Delta v)$}
\newcommand{\xvr}{$\xi(\Delta v, \Delta r)$}
\newcommand{\civ}{\ion{C}{iv}}
\newcommand{\kms}{$\,{\rm km}\,{\rm s}^{-1}$}
\newcommand{\logN}{$\log (N /\rm{cm}^{-2})$}

\begin{document}

\title{The multiple coherence scales of \civ\ at cosmic noon} 

\author{
H. Cortés-Muñoz\inst{1}
\and S. Lopez\inst{1}
\and N. Tejos\inst{2}
\and J.-K. Krogager\inst{3,4}
\and D. Zamora\inst{1}
\and R. Cuellar\inst{1}
\and P. Anshul\inst{1}
\and F. Urbina\inst{5}
\and A. Afruni\inst{6,7}
}

\institute{Departamento de Astronomía, Universidad de Chile, Casilla 36-D, Santiago, Chile. \email{hcortes@das.uchile.cl}
\and Instituto de Física, Pontiﬁcia Universidad Católica de Valparaíso, Casilla 4059 Valparaíso, Chile
\and French-Chilean Laboratory for Astronomy, IRL 3386, CNRS and U. de Chile, Casilla 36-D, Santiago, Chile
\and Centre de Recherche Astrophysique de Lyon, Université de Lyon 1, ENS-Lyon, UMR5574, 9 Av Charles André, 69230 Saint-Genis-Laval, France
\and Kapteyn Astronomical Institute, University of Groningen, Landleven 12, 9747 AD Groningen, The Netherlands
\and Dipartimento di Fisica e Astronomia, Università di Firenze, Via G. Sansone 1, 50019 Sesto Fiorentino, Firenze, Italy.
\and INAF - Osservatorio Astrofisico di Arcetri, Largo E. Fermi 5, Firenze, I-50125, Italy.
}

\date{Received November 15 2025 / Accepted January 21 2026}

\abstract{
The spatial and kinematic structure of the circumgalactic medium remains poorly constrained observationally. We computed the clustering of \civ\ absorption systems at cosmic noon using quasar pairs. We analyzed VLT/UVES and Keck/HIRES high-resolution spectra ($R \approx 45\,000$) of a sample of eight projected and four lensed quasar pairs that probe transverse separations, \sep, from subkiloparsec to a few megaparsec over the redshift range $1.6 \lesssim z \lesssim 3.3$. We detected and fit Voigt profiles to a total of 141 \civ\ systems, corresponding to 620 velocity components at all quasar lines of sight. We computed the two-point correlation function of \civ, $\xi(\Delta v, \Delta r)$, where $\Delta v$ is the velocity difference between components at all available scales. We found a strong dependence of $\xi(\Delta r)$ on \sep\ at all velocities. $\xi(\Delta r)$ reaches a sharp peak at the smallest scales we analyzed, $\Delta r\approx 0.1$ kpc, decreases steadily up to $\Delta r\approx 5$ kpc, and remains flat up to $\Delta r\approx 500$ kpc, where it again begins to decrease. By fitting power laws to the projected transverse correlation function \Xir, we inferred two coherence lengths. The first is $r_1 = 654^{+100}_{-87}$ kpc, which we interpret as a representative size for the \ion{C}{iv} enriched regions at $z\approx 2$, and the second is $r_2 = 4.70^{+1.60}_{-1.19}$ kpc for the individual \ion{C}{iv}-bearing clouds. When we instead projected this in $\Delta r$, we found amplitudes of $\xi(\Delta v)$ that were consistent with those in previous works that used quasars and extended background sources. Our results suggest that \civ\ might be a good tracer of not only the small internal structure of the circumgalactic medium, but also of the way in which galaxies cluster at cosmic noon. 
}


\maketitle

\nolinenumbers

\section{Introduction} \label{introduction}

The formation and evolution of galaxies is thought to leave an indelible mark on the spatial and kinematic structure of the circumgalactic medium (CGM; \citealp{Tumlinson17, Peroux20, Faucher23}). This is due to the baryon cycle, in which the gas flows in and out of galaxies through the CGM \citep{Keres05, Oppenheimer06, Faucher11, Nelson16}. Since the CGM is diffuse 
and usually too faint for direct observations, this gas is typically studied in absorption toward bright background sources such as quasars (e.g., \citealp{DOdorico10, Cooksey13, Kim13, Lehner18, Churchill20}), galaxies (e.g., \citealp{Steidel10, Rubin14, Bordoloi17, Cooke15, Chen20}), and the afterglow of gamma-ray bursts (e.g., \citealp{Prochter2006, Tejos2007,Tejos2009, Vergani09, Bolmer18, Saccardi23, Krogager24}). 

The use of single lines of sight prevents an unambiguous determination of the spatial structure and clustering of the gas, however.  An alternative is the use of the close parallel lines of sight provided by multiple quasars in general \citep{Hennawi06, Coppolani06, Martin10, Tejos14, Mintz22, Urbano23}, lensed quasars \citep{Rauch99, Lopez1999,Lopez05,Lopez07, Chen14, Zahedy16, Krogager18, Cristiani24, Dutta24}, and lensed galaxies \citep{Lopez18, Lopez20, Mortensen21, Tejos21, Bordoloi22, FernandezFigueroa22, Afruni23, Lopez24}. Remarkably, multiple lines of sight considerably expand the information provided by single lines of sight, even though multiple lines of sight are rare. This promises unique clues on the extent and coherence of the absorbers.

In this work, we concentrate on the spatial structure of \civ\ absorption systems that are signposted through the \ion{C}{iv} $\lambda\lambda$1548,1550 doublet, which is an easy-to-find and representative tracer of the cool and warm CGM at $z \gtrsim 1.5$~\citep[e.g., ][]{Schaye03, Songaila05, DOdorico10, DOdorico13, Cooksey13, Boksenberg15, Hasan20}. Measurements using transverse~\citep[e.g., ][]{DOdorico06,Martin10, Maitra19,Mintz22} and line-of-sight~\citep[e.g., ][]{Scannapieco06,Boksenberg15,Welsh2025} correlations show that the clustering amplitude of strong \civ\ absorbers on megaparsec scales is comparable to galaxy clustering. Since the \civ-galaxy cross correlation is also significant \citep{Adelberger05b,Banerjee23}, the picture that emerges is consistent with \civ\  predominantly arising in enriched gas associated with galaxy halos and their surrounding environments. However, studies on the clustering of the CGM are mostly focused on large scales~\citep[e.g., ][]{DOdorico06,Martin10, Maitra19,Mintz22}, and only very few ventured into the small-scale clustering of the CGM (e.g., \citealt{Rauch01,Lopez24}). We present the largest archival sample of echelle spectra of closely separated quasar pairs and use it to measure the \ion{C}{iv} two-point autocorrelation function over transverse scales from the subkiloparsec and up to a few megaparsec. This allows us to estimate the scale of the CGM and the sizes of individual clouds of \ion{C}{iv}. We assume a $\Lambda$CDM cosmology with parameters $H_0=70\,{\rm km}\, {\rm s}^{-1}\, {\rm Mpc}^{-1}$, $\Omega_m=0.3$, and $\Omega_\Lambda=0.7$ throughout. Every distance measurement is comoving, unless stated otherwise.

\section{Data}\label{data}
\subsection{Sample construction} 

\begin{table*}
    \caption{Lensed and projected quasars pairs.}
    \centering
    \begin{tabular}{cccccccc}
    \hline
    \hline
    Type & $z_{lens}$	 & Quasar (A) &  $z_{quasar}$ (A) & Quasar (B) & $z_{quasar}$ (B)& Separation & Reference	\\
                 &              &        &				&		&	 & (")  		& \\ 
    \hline

    Lensed       & 0.490    & Q0142-100 (A)		&  2.720   & Q0142-100 (B)           & 2.720      & 2.24     & (1)\\
    Lensed       & 0.770    & RX J0911.4+0551 (A) &  2.800   & RX J0911.4+0551 (B)     & 2.800      & 2.47     & (2)\\
    Lensed       & 0.550    & J1029+2623 (A)		&  2.197   & J1029+2623 (B)          & 2.197      & 22.50    & (3)\\
    Lensed       & 0.730    & HE 1104-1805 (A)	&  2.320   & HE 1104-1805 (B)        & 2.320      & 3.19     & (2)\\
    Projected    &          & J000852-290043 		&  2.645   &  J000857-290126 		& 2.607      & 78.48      & (4)\\
    Projected    &          & J030640-301031 		&  2.096   &  J030643-301107 		& 2.130      & 51.66      & (4)\\
    Projected    &          & J031006-192124 		&  2.122   &  J031009-192207 		& 2.144      & 60.47      & (4)\\
    Projected    &          & J143229-010614 		&  2.087   &  J143229-010616 		& 2.076      & 4.12       & (4)\\
    Projected    &          & J214222-441929 		&  3.230   &  J214225-442018 		& 3.220      & 62.14      & (4)\\
    Projected    &          & J223948-294748 		&  2.121   &  J223951-294836 		& 2.068      & 63.60      & (4)\\
    Projected    &          & J234625+124743 		&  2.515   &  J234628+124858 		& 2.568      & 85.04      & (4)\\
    Projected    &          & J234819+005717 		&  2.163   &  J234819+005721 		& 2.145      & 7.03       & (4)\\ 
    \hline
    \end{tabular}
    \tablebib{(1) \cite{OMeara15, OMeara17}; (2) \cite{Lopez05, Lopez07}; (3) PID: 092.B-0512(A) (PI Saez); (4) \cite{Murphy19}}
    \label{table:qso}
\end{table*}

To build a statistically significant sample of quasar pairs with high-resolution spectroscopy that was as complete as possible, we searched for quasar spectra in two primary databases: the Keck Observatory Database of Ionized Absorption toward Quasars (KODIAQ; \citealt{OMeara15, OMeara17}), using High Resolution Echelle Spectrometer (HIRES; \citealt{Vogt94}) spectra from the Keck Observatory, and the Spectral Quasar Absorption Database (SQUAD; \citealt{Murphy19}), using Ultraviolet and Visual Echelle Spectrograph (UVES; \citealt{Dekker00}) spectra from the Very Large Telescope (VLT) observatory. We used the tool \texttt{TOPCAT} (\citealt{Taylor20}) to find suitable on-sky projected pairs. We limited the search to pairs with maximum transverse physical separations shorter than 1 Mpc (this translates into several comoving megaparsec at these redshifts; see Section \ref{sep}) and emission redshifts $z_{em} > 2$ to ensure coverage of \civ. The search delivered nine quasar pairs, one of which is a lensed quasar (see Table \ref{table:qso}). These spectra were reduced and the continuum was normalized on their respective databases. We complemented the sample with additional high-resolution spectra of three lensed pairs found in the literature: J1029+2623 (PID: 092.B-0512(A)), HE 1104-1805, and RX J0911.4+0551 \citep{Lopez05, Lopez07}. All archival spectra were reduced and normalized in a standard fashion. The exception was J1029+2623. For this case, its reduced spectra were extracted from the archive and were then combined to a common array by interpolating their wavelengths and averaging each flux pixel weighting by the inverse variance of each spectrum. The continuum normalization for this pair was carried out independently for the two lines of sight by spline fitting using the code \texttt{LINETOOLS} \citep{Prochaska16}. The final sample consisted of the resolved echelle spectra of four lensed and eight projected quasar pairs and is summarized in Table \ref{table:qso}. The emission redshifts lie in the range $2.08 \le z \le 3.22$. The resolving power is around $R = \lambda/\Delta\lambda \approx 45\,000$ ($\approx 6.6$ \kms), and the typical wavelength range is $3000-10000$ \r{A}, with variable coverage depending on the target. The signal-to-noise ratios range between 5 and 100. 

\subsection{Redshift path}

\begin{figure}
  \centering
  \includegraphics[width=1\columnwidth]{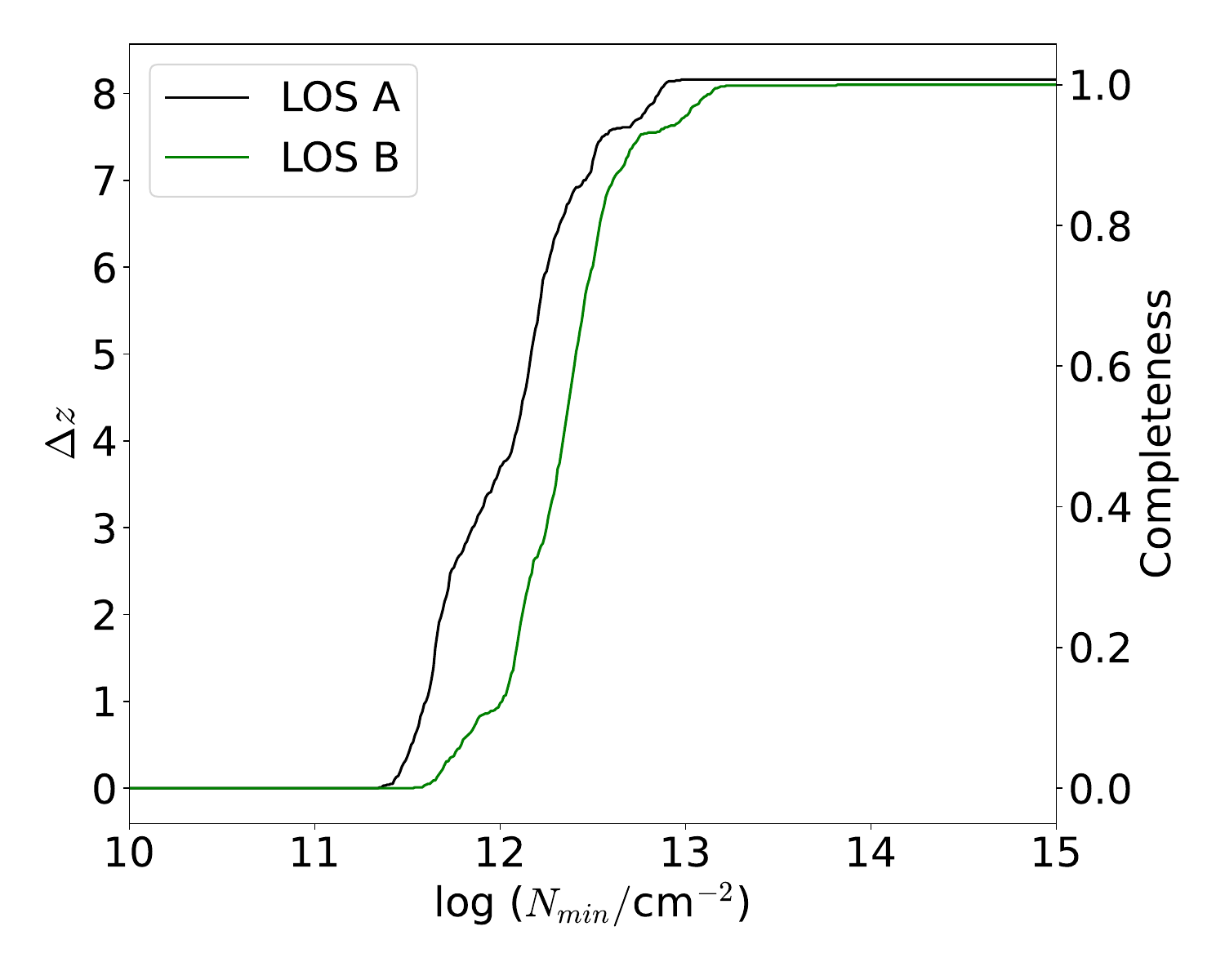}
  \caption{Redshift path ($\Delta z$) and completeness for the line-of-sight A (B) in black (green) as a function of the column density threshold ($N_{min}$).}
\label{zpath}
\end{figure}

The signal-to-noise at a given wavelength on the line-of-sight A (the brighter spectrum) and B (the fainter spectrum) are usually different. To evaluate these differences, we computed the redshift path density for a \civ\ detection,
\begin{equation}
    g(N_{min}, z) = \sum_{n=1}^{N_{LOS}} H(z - z_n^{min})H(z_n^{max}-z)H(N_{min}-N_{lim}(z))~,
\end{equation}
where $H$ is the Heaviside step function, $z_n^{min}$ and $z_n^{max}$ are the redshift ranges of the line-of-sight number $n$, where we searched for \civ\, between the Ly$\alpha$ and \civ\ emission lines of the quasar, $N_{min}$ is the column density threshold, and $N_{lim}$ is the minimum column density observable at redshift $z$. $N_{lim}$ is computed at the $3\sigma$ level as \citep{Savage91}
\begin{equation}\label{N_lim}
N_{lim} = 1.13\times10^{20} \dfrac{W_{lim}}{\lambda_0^2f}~\text{cm}^{-2}~,
\end{equation}
where $f$ is the oscillator strength of the transition with the wavelength $\lambda_0 = 1548.2040$, and $W_{lim} = {3 \times \text{FWHM}}/{\langle S/N \rangle  (1+z)}$. Here, $\langle S/N \rangle$ corresponds to the average signal-to-noise ratio around redshift $z$, and FWHM $= \lambda/R$. Finally, by integrating $g(N_{min}, z)$ over redshift, we obtained the redshift path, $\Delta z = \int_{z_{min}}^{z_{max}} g(N_{min}, z) dz$. Fig. \ref{zpath} shows $\Delta z(N_{min})$ for each line of sight. The $\Delta z(N_{min})$ can be associated with a completeness level by comparing it with the total $\Delta z$. A 90\% completeness for a \civ\ detection is reached at \logN\ = 12.51 for lines-of-sight A and at \logN\ = 12.69 for lines-of-sight B. In this way, we found a total redshift path (as given by an $N_{min}$ that gives a 100\% completeness) of $\Delta z \approx 8.16$ on lines-of-sight A and $\Delta z \approx 8.1$ on lines-of-sight B. The small difference in $\Delta z$ between the lines of sight is due to very small differences in the wavelength coverage. We address this difference in $\Delta z$ with the construction of random absorption catalogs to compute the two-point autocorrelation function in Section \ref{TPCF}.

\section{Absorption line analysis}

\subsection{Absorption line identification and fitting} \label{VP}

\begin{figure*}
  \centering
  \includegraphics[width=1\columnwidth]{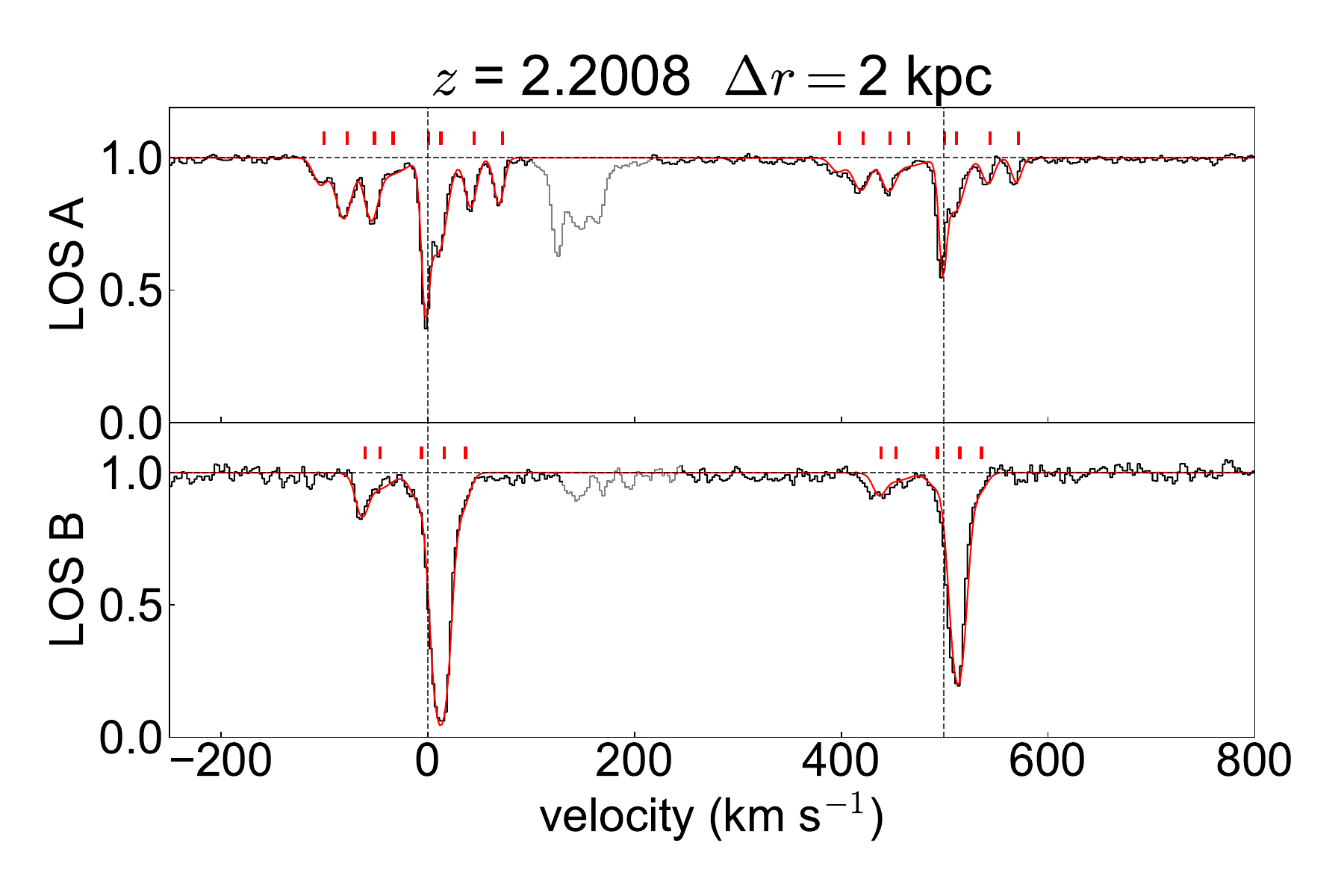}
  \includegraphics[width=1\columnwidth]{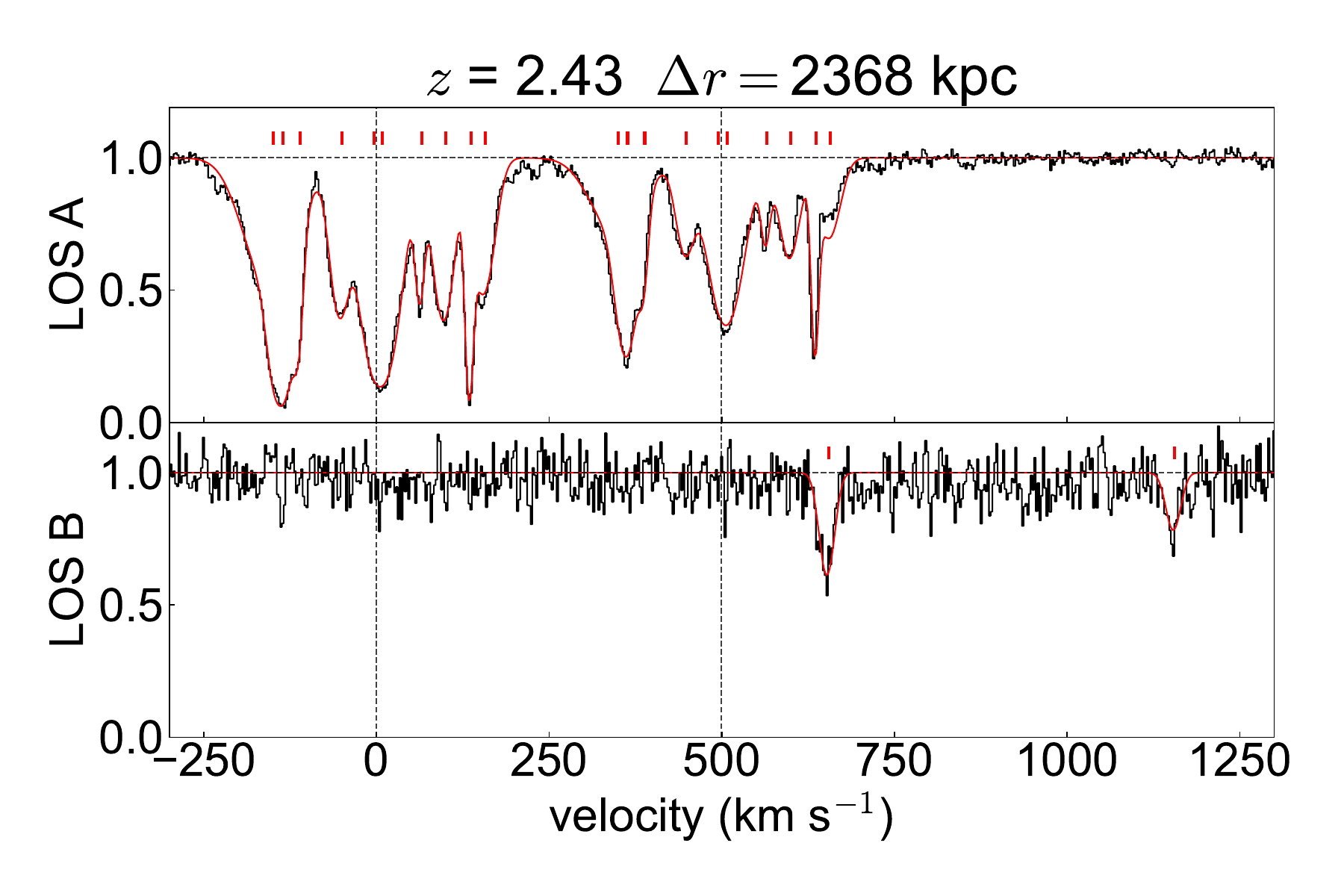}
  \caption{Examples of \ion{C}{iv} absorption systems and their Voigt profiles fits. Left panel: Absorption system at $z_{abs} = 2.2008$ toward the lensed quasar HE 1104-1805. Right panel: Absorption system at $z_{abs} = 2.43$ toward J234628+124858 and J234625+124743. The top (bottom) panels correspond to line-of-sight A (B). The black histogram corresponds to the normalized flux of the quasars, and the red lines correspond to the Voigt profile fit of the system. The velocity zeropoint was chosen by eye around the 1548 \r{A} \ion{C}{iv} line.}
\label{sys}
\end{figure*}

To identify \civ\ absorption systems, we automatically searched for peaks in the optical depth in the wavelength range between Ly$\alpha$ and \civ\ emission lines and selected those by eye that were compatible with the doublet line separation and the doublet ratio. This approach identified 73 systems for lines-of-sight A and 68 for lines-of-sight B. To accurately measure the redshifts of the individual \civ\ velocity components on each absorption system, we fit Voigt profiles to each system using the Python package \texttt{VoigtFit} (\citealt{Krogager18_VF}). Before the fit, we masked each region for bad pixels. When interlopers overlapped with a \civ\ absorption system, they were fit together. The fits were performed on each line of sight independently. In a first approach, we considered a fit successful when a \texttt{VoigtFit} model converged for the whole system. Moreover, we computed a component-by-component badness parameter \citep{Churchill20}, 
\begin{equation}
    \text{badness} = \Bigg{[}\Bigg{(}\dfrac{\sigma_N}{N}\Bigg{)}^2 + \Bigg{(}\dfrac{\sigma_b}{b}\Bigg{)}^2 + \Bigg{(}\dfrac{\sigma_z}{z}\Bigg{)}^2\Bigg{]}^{1/2},
\end{equation}
\noindent where $\sigma_N$, $\sigma_b$, and $\sigma_z$, are the uncertainties determined by \texttt{VoigtFit} for column density ($N$), Doppler parameter ($b$), and redshift ($z$), respectively. Individual velocity components with badness $>1.5$ were refit with different initial parameters. This robustness check helped us to find more physically motivated components and avoided unreasonably large or small results for the fit parameters without much intervention. The final accepted fits were still a human decision, however. Considering all pairs, we found 327 \civ\ velocity components in lines-of-sight A and 293 in lines-of-sight B. The redshift range of this sample is $1.51 < z < 3.23$. Fig \ref{sys} shows two examples of absorption systems, along with their Voigt profile fits, at two extreme comoving transverse separations, \sep. Qualitatively, this comparison suggests that smaller \sep\ probes more coherent structures than larger \sep.

\subsection{Column density distribution}

\begin{figure}
  \centering
  \includegraphics[width=1\columnwidth]{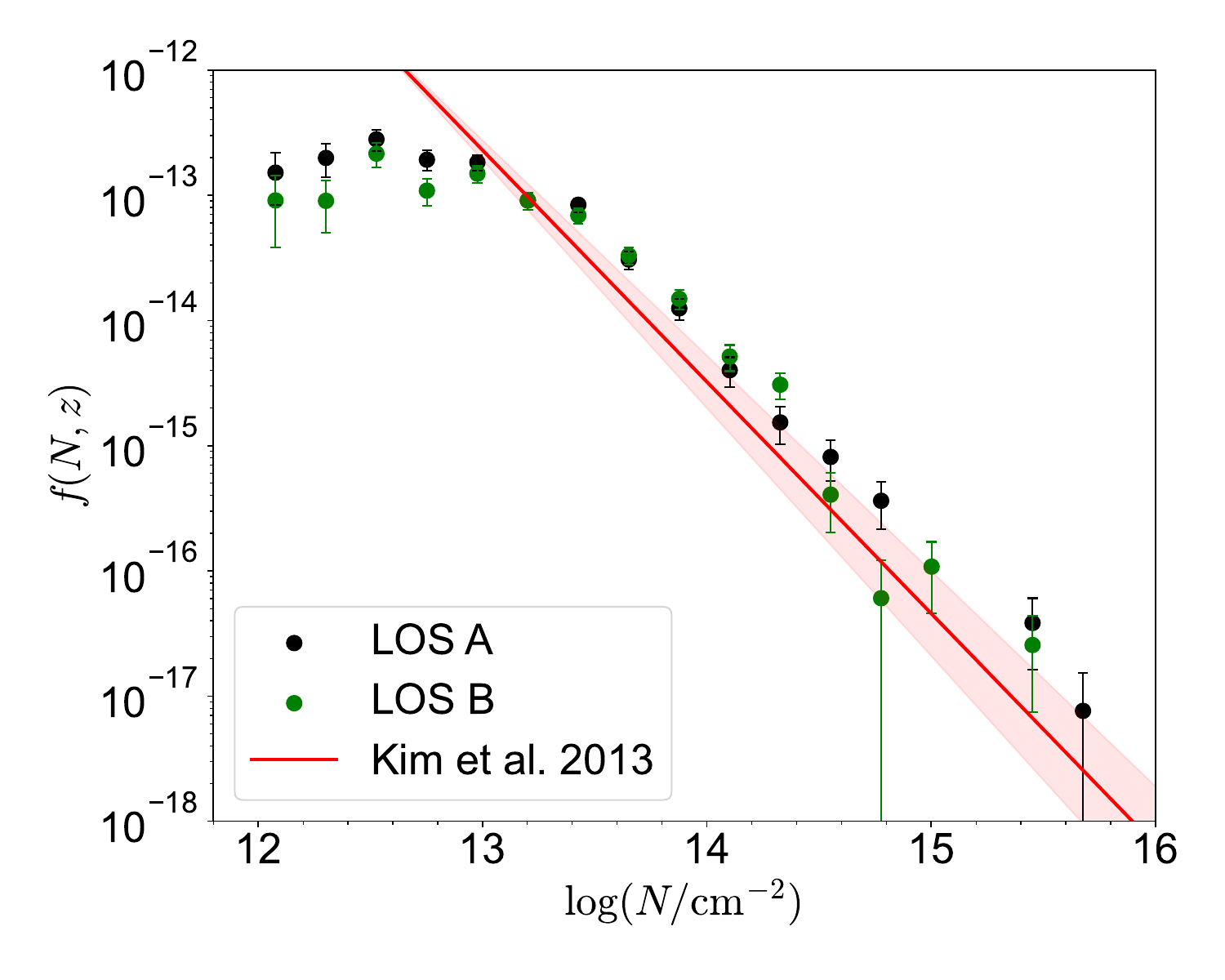}
  \caption{\ion{C}{iv} column density distribution for lines-of-sight A in black and lines-of-sight B in green. The solid pink line corresponds to the power-law fit by \citet{Kim13} to data of 18 VLT/UVES single-quasar sightlines, and the shaded area is the 1$\sigma$ uncertainty in the fit.}
\label{fNz}
\end{figure}

Another way to assess the completeness of our sample is the column density distribution, $f(N,z)$. We calculated $f(N, z)$ following Eqs. (6) and (7) of \cite{Zafar13}. In Fig. \ref{fNz}, we show $f(N, z)$ independently for lines-of-sight A and B. This distribution can be fit by a power law, $f(N, z) = C \times N^{-\delta}$, where $C$ is a normalization constant, and $\delta$ is the slope of the power law. We obtained the best-fit parameters $\log (C) = 9.00 \pm 0.78$, and $\delta = 1.66 \pm 0.06$ for lines-of-sight A, and for lines-of-sight B, we found $\log (C) = 10.58 \pm 1.81$, and $\delta = 1.77 \pm 0.13$. We limited these fits to $N > 10^{13}$ cm$^{-2}$, where A and B have similar completeness levels. Our fits agree well with previous results by \citet{Kim13} at $1.9 < z < 3.2$ ($\delta = 1.85 \pm 0.13$) and \citet{DOdorico10} at $1.6 < z < 3.6$ ($\delta = 1.71 \pm 0.07$), who used a similar redshift path and instrumental resolution. This agreement is a strong diagnostic for the completeness of the survey and supports our confidence in the Voigt profile fitting.

\subsection{Transverse line-of-sight separations} \label{sep}

\begin{figure}
  \centering
  \includegraphics[width=1\columnwidth]{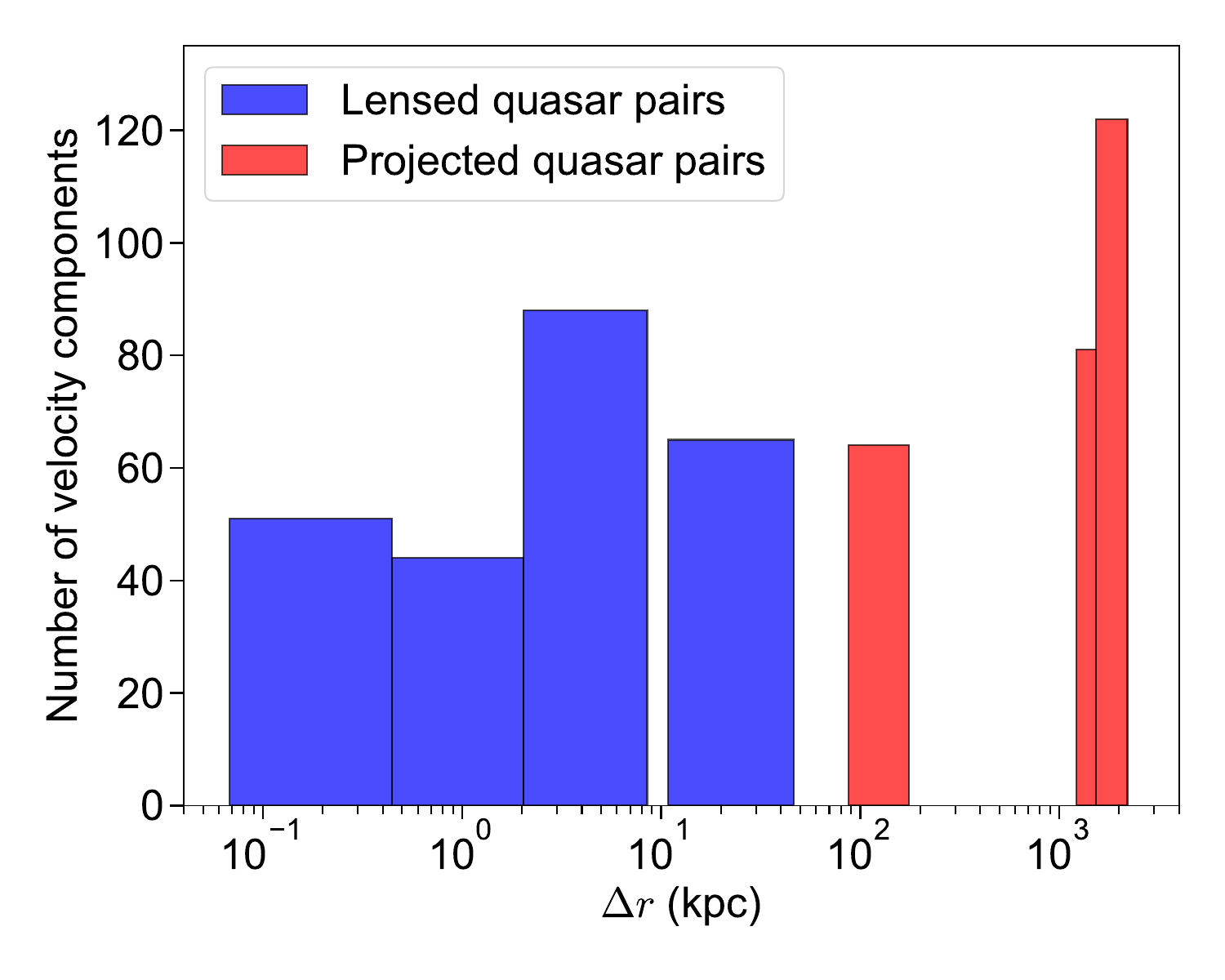}
  \caption{Distribution of the \civ\ velocity components. The blue histogram corresponds to velocity components found on lensed quasars pairs, and the red histogram corresponds to velocity components found on projected quasar pairs.}
\label{vc_sep}
\end{figure}

For the projected pair case, we computed $\Delta r$ between lines of sight following \cite{Hogg99}. For the lensed quasars, we calculated the separations by assuming that the lens is produced by a point-like source, following \cite{Krogager18} 

\begin{equation}
    \Delta r (z) = 
    \begin{cases}
        \theta \dfrac{D_{AS}D_{OL}}{D_{LS}} & \text{lensed quasars} \\
        \theta \cdot D_{OA} & \text{projected quasar pairs,}
    \end{cases}
\end{equation}

\noindent where $\theta$ is the angular separation on the sky of the quasar images, while the angular diameter distances are $D_{AS}$ from the absorber to the quasar, $D_{OL}$ from the observer to the lens, $D_{LS}$ from the lens to the quasar, and $D_{OA}$ from the observer to the absorber. The resulting range of \sep\ for the sample is $20 \text{ pc} \lesssim \Delta r \lesssim 2.4 \text{ Mpc} $. Fig. \ref{vc_sep} shows the resulting distribution of \sep. The space between some bins is produced by gaps in the survey in which no quasar pairs were found. We chose this binning to account for roughly a similar number of velocity components in each bin without splitting the redshift path of a quasar line of sight. This binning is used throughout Section \ref{TCA}.

\section{Transverse correlation analysis} \label{TCA}

\subsection{Two-point correlation function} \label{TPCF}

\begin{figure}
  \centering
  \includegraphics[width=1\columnwidth]{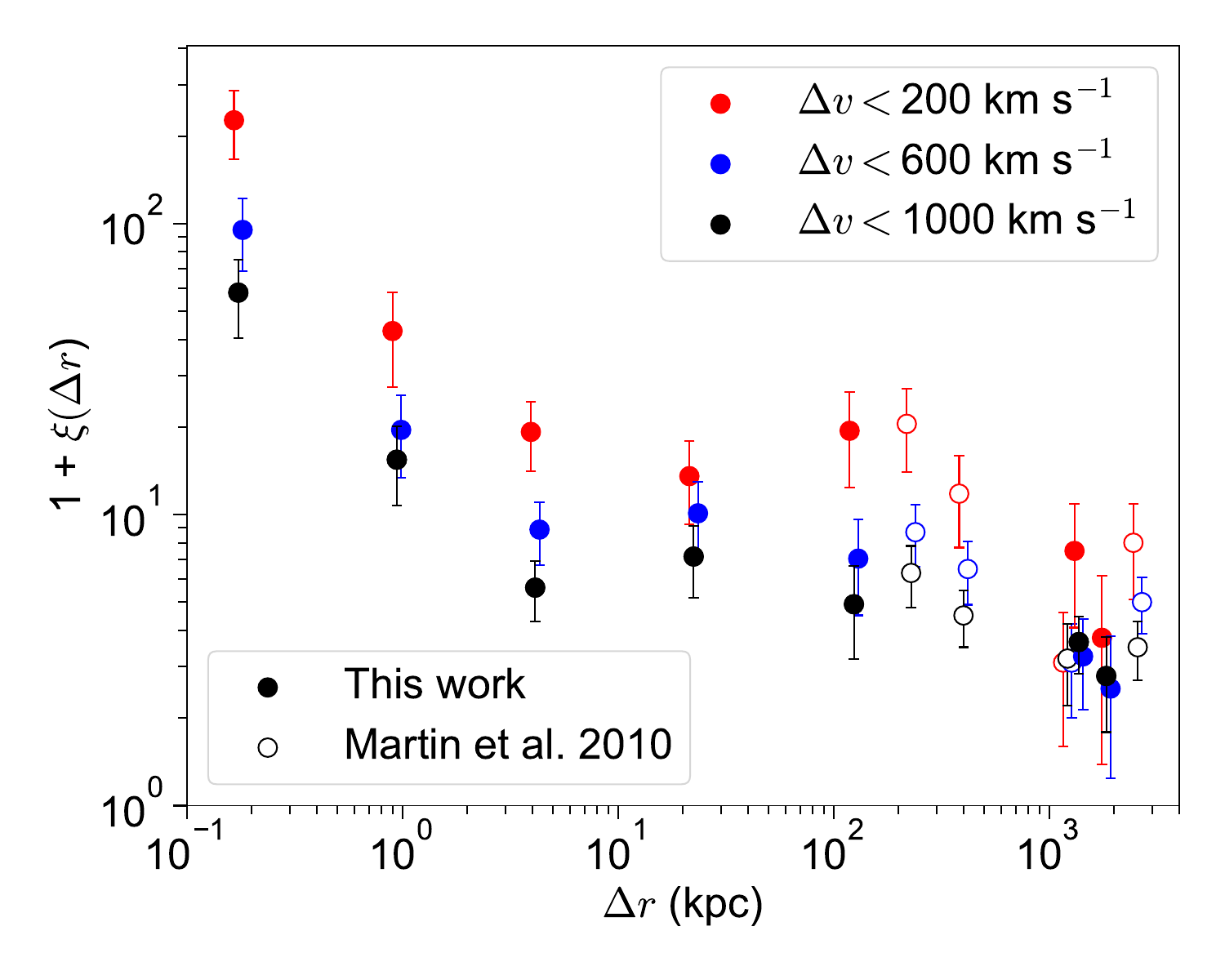}
  \caption{Two-point correlation function ($\xi(\Delta r)$) of \ion{C}{iv} as a function of the line-of-sight separation for three different velocity bins. The red and blue points are slightly shifted for clarity from their identical position to the blue points. The open circle data points correspond to the measurements obtained by \cite{Martin10}.}
\label{acf_r_civ}
\end{figure}

In order to study how \civ\ clusters at different scales, we defined the two-point correlation function, $\xi(\Delta r, \Delta v)$, as the probability excess with respect to a randomly distributed sample,of finding a pair of velocity components with redshift $z_A$ on lines-of-sight A and $z_B$ on B with a velocity difference $\Delta v = c |z_A - z_B|(1 + z_{AB})^{-1}$, where $z_{AB}$ is the average redshift, and a line-of-sight separation $\Delta r$. To compute $\xi(\Delta r, \Delta v)$, we followed implementations similar to those by \cite{Martin10} and \cite{Tejos14} of the generalized \cite{Landy93} estimator, in our notation:

\begin{equation}
\xi(\Delta v, \Delta r) = \frac{D_A D_B/n_{DD} - D_A R_B/n_{DR} - D_B R_A/n_{DR}}{R_A R_B/n_{RR}} +1,
\end{equation}

\noindent where $D_A D_B$ is the number of data-data pairs, $R_A R_B$ is the number of random-random pairs, and $D_A R_B$ and $D_B R_A$ are the data-random pairs. These pairs were computed by creating every possible combination of velocity components between lines-of-sight A and B. The data-random pairs were built by pairing the data catalog from one line of sight with a random catalog generated with the other. These quantities were normalized by their respective normalization factors, 

\begin{equation} \label{eq:norm}
\begin{tabular}{l}
    $n_{DD} = N_A N_B$~, \\
    $n_{DR} = \alpha N_A N_B$~, and\\
    $n_{RR} = \alpha^2 N_A N_B$~;
\end{tabular}
\end{equation}

\noindent where $N_A$ and $N_B$ are the total number of absorption lines found on lines-of-sight A and B, respectively, and $\alpha$ is a positive integer used to build the random catalogs, which we describe below. To create these pairs of velocity components, we only considered column densities above $90\%$ completeness in line-of-sight B, and this same limit was applied on line-of-sight A. In total, this corresponds to 300 components in A and 249 components in B ($\approx88\%$ of the total sample).

The creation of random catalogs is a delicate point when working with two-point correlation functions, as it largely depends on the geometry and type of objects involved. We followed the approach of \cite{Tejos14} and created $\alpha = 100$ copies of each measured parameter set $(z, N)$ by randomizing their redshifts. Each random redshift was drawn uniformly from any position on the quasar line of sight where the corresponding $N$ could have been measured, depending on the spectral coverage and the column density limit given the signal-to-noise ratio. This limiting column density was measured following Eq. \ref{N_lim}. This way of randomizing the redshifts aids in reducing any sort of detection bias because the spectra naturally have a different signal-to-noise at different wavelengths.

The \cite{Landy93} estimator is typically preferred because it minimizes the Poisson variance ($\Delta^2_{DD} (\xi) = (1+\xi)/DD$). When only uncertainties due to Poisson variance are assumed, however, as discussed by \cite{Mo92} and \cite{Tejos14}, the real uncertainty might be underpredicted. To remedy this potential issue, we adopted a more conservative approach by estimating our uncertainties ($\Delta^2_{BS}$) via bootstrapping and computed our correlation measurements $N_{BS} = 100$ times with a random sample (with repetition) of $N_A$ velocity components on line-of-sight A and $N_B$ on line-of-sight B. For every bin,
\begin{equation}
    \Delta^2_{\text{BS}}(\xi) = \dfrac{1}{N_{BS}} \sum_i^{N_{BS}} (\xi_i - \bar{\xi})^2~,
\end{equation}
where $\bar\xi$ is the mean over all bootstrap measurements $\xi_i$.   
Finally, to compute $\xi$, each type of pairs ($D_AD_B$, $D_AR_B$, $D_BR_A$, $R_AR_B$) was summed, 
\begin{equation}
    D_AD_B(\Delta v, \Delta r) = \sum_i^{N_{LOS}} D_AD_B(\Delta v, \Delta r)_i.
\end{equation}

In Fig. \ref{acf_r_civ} we show $1+\xi(\Delta r)$ binned for three values of maximum $\Delta v$ as a function of \sep. The bins in separation were chosen so that each bin had roughly the same number of pairs when possible (the same binning is used in Fig. \ref{vc_sep}). The maximum $\Delta v$ (200, 600, and 1000 \kms) was chosen to compare more directly with \cite{Martin10} (open circles), who analyzed data of 29 projected quasar pairs at a medium spectral resolution. All of these pairs are different from the 12 pairs we found on our sample, which makes our results completely independent. The figure shows that the amplitude of \xir generally increases with decreasing separation, although at scales between $\approx 5$ kpc and $\approx 500$ kpc, $\xi$ seems to be constant in three bins when it is compared with the smaller bin by \cite{Martin10}. 

\subsection{Projected transverse correlations}\label{tcf}

\begin{figure}
  \centering
  \includegraphics[width=1\columnwidth]{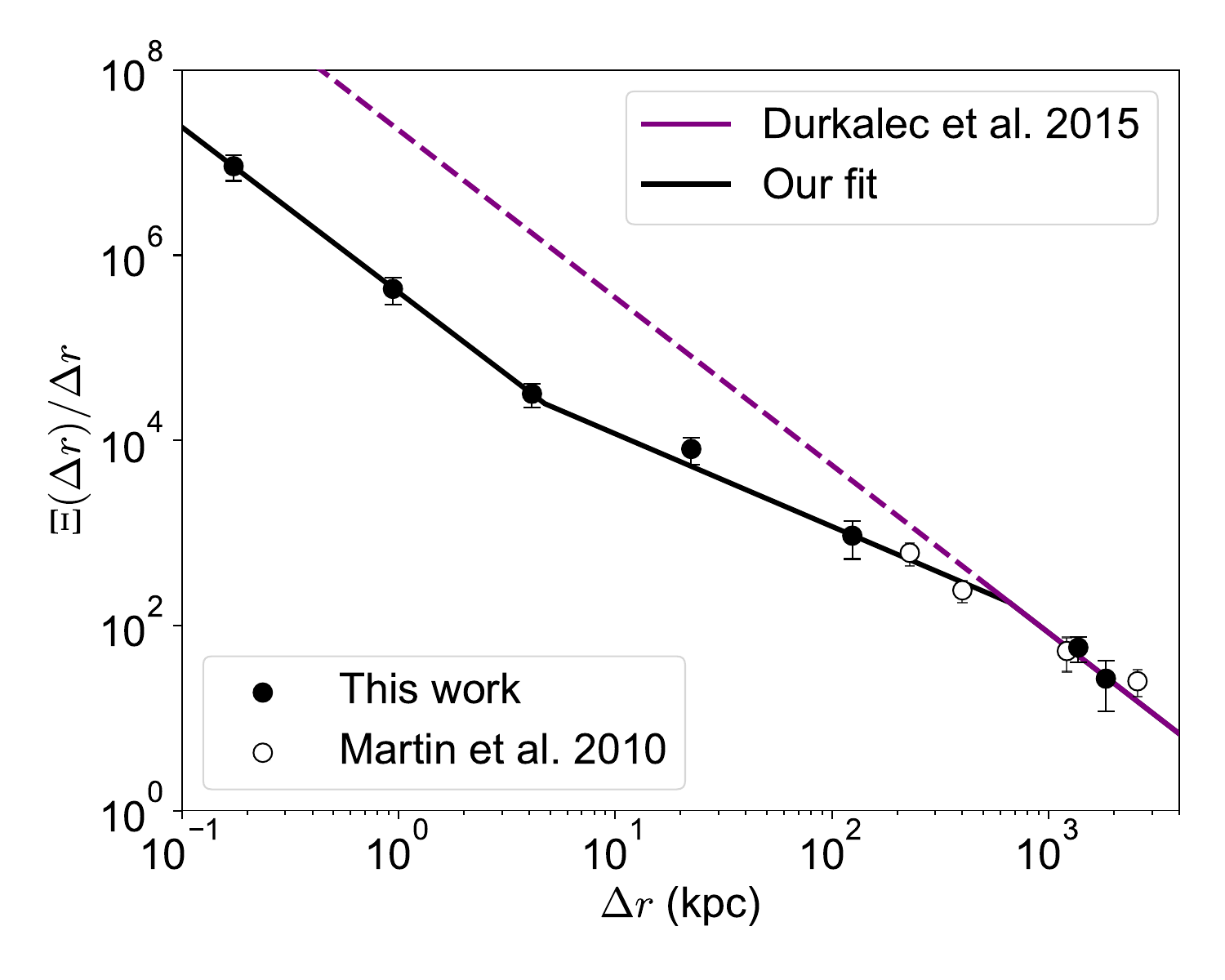}
  \caption{Projected transverse correlation function ($\Xi (\Delta r)$) for \ion{C}{iv}. The bins are the same as in Fig. \ref{acf_r_civ}. The open circle data points correspond to the measurements obtained by \cite{Martin10}, using the same velocity threshold and estimator. The solid purple line shows a galaxy-galaxy autocorrelation function result \citep{Durkalec15}, with parameters $r_0 = 5.64^{+0.69}_{-0.77}$ Mpc and $\gamma = 1.81^{+0.02}_{-0.06}$, and the dashed purple line is its extrapolation to small scales. The solid black line corresponds to the model of Eq. \ref{eq:break2}, which described further in the main text.
  }
  
\label{tcf_civ}
\end{figure}

In order to compare our measurements with galaxy autocorrelation studies and properly fit $\xi(\Delta r)$, we computed the projected transverse correlation function (PTCF) \citep{Davis83}. The PTCF is the integral of $\xi(\Delta v, \Delta r)$ over the distance in redshift space ($\Delta \Pi$), which is obtained by assuming the observed $\Delta v$ is produced by Hubble flow, and therefore, $\Delta\Pi(\Delta v, z) = (1+z) \Delta v / H(z)$. Then, the PTCF is computed as

\begin{equation}
    \Xi(\Delta r, \Delta\Pi_{max}) = 2 \int_0^{\Delta\Pi_{max}} \xi(\Delta \Pi, \Delta r)\text{d}\Delta \Pi.
\end{equation}

Additionally, this can be approximated because of the cylindrical symmetry of the survey following \cite{Martin10} and \cite{Tejos14} by

\begin{equation}
\Xi(\Delta r, \Delta\Pi_{max}) \approx 2\Delta \Pi_{max} \xi(\Delta \Pi_{max}, \Delta r).
\end{equation}

Moreover, \cite{Davis83} showed that this function can be fit using a power law with a slope $\gamma$ by assuming that the correlations in real space take the form $\xi(\Delta r) = (r_0/\Delta r)^\gamma$. Then 

\begin{equation} \label{eq:fit}
\Xi(\Delta r) = A(r_0, \gamma)r^{1-\gamma},
\end{equation}

\noindent where $A(r_0, \gamma) = r_0^\gamma \Gamma(1/2)\Gamma[(\gamma-1)/2]/\Gamma(\gamma/2)$, and $\Gamma$ is the gamma function. It is important to note that this assumes a $\gamma > 1$ and $\Pi_{max} \rightarrow \infty$. Then, a fit of $\Xi(\Delta r)$ intermediately provides a value for $r_0$ and $\gamma$. In Fig. \ref{tcf_civ} we show $\Xi(\Delta r)$ for the same velocity threshold bins as in Fig. \ref{acf_r_civ}. We chose the larger threshold because the condition for Eq. \ref{eq:fit} is that $\Pi_{max} \rightarrow \infty$.

We then fit $\Xi(\Delta r)$ in three different ways. First, we fit a single power law (Eq. \ref{eq:fit}) with best-fit parameters $r_0 = 4.63 \pm 1.18$ Mpc and $\gamma = 1.18 \pm 0.04$ (see Fig. \ref{tcf_civ1}). While this fit works well at higher $\Delta r$, it underpredicts the smaller separation bin, and it does not consider the relation between \civ\ and galaxies. With this in mind, we considered the prediction from \cite{Scannapieco06} that $\xi$ should break and flatten somewhere at small separations. 

Following \cite{Martin10}, we modified our fit to \Xir\ to add a break to the power law of the form

\begin{equation} \label{eq:break1}
    \xi (\Delta r) = 
    \begin{cases}
        (r_0/r_1)^\gamma, & \Delta r<r_1 \\
        (r_0/\Delta r)^\gamma, & \Delta r \ge r_1~,
    \end{cases}
\end{equation}

\noindent where the first regime delimited by $r_1$ is seen as a constant correlation $\xi(\Delta r)$ (which when normalized is seen as $\Xi(\Delta r)/\Delta r \propto 1/\Delta r$). Following \cite{Martin10}, we adopted the $r_0$ and $\gamma$ parameters form the galaxy autocorrelation function; this can be assumed when we consider that the correlation amplitudes of \ion{C}{iv} and those of the galaxies are the same at scales larger than the sizes of \ion{C}{iv} enriched regions. We chose the result of \cite{Durkalec15} with parameters $r_0 = 5.64^{+0.69}_{-0.77}$ Mpc and $\gamma = 1.81^{+0.02}_{-0.06}$ because they studied a sizable sample of 1556 galaxies in a similar redshift range $2.0 < z < 2.9$. In this way, the only free parameter for this model is $r_1$, for which we found a best fit at $r_1 = 597 \pm 130$ kpc when we considered the points from \cite{Martin10} (see Fig. \ref{tcf_civ2}). The flat $\xi(\Delta r)$ prediction still shows an even larger offset at small \sep\ when compared with the single-power law model, however. This is also noticeable in the increasing correlations at very small scales ($\Delta r < 10$ kpc) and is best seen in Fig. \ref{acf_r_civ}. 

Taking this into consideration, we modified Eq. \ref{eq:break1} to add a second break on $\xi(\Delta r)$ to account for a third regime at the smallest separations,

\begin{equation} \label{eq:break2}
    \xi (\Delta r) = 
    \begin{cases}
        (r_3/\Delta r)^{\gamma_1}, & \Delta r \le r_2 \\
        (r_0/r_1)^\gamma, & r_2 < \Delta r < r_1 \\
        (r_0/\Delta r)^\gamma, & \Delta r \ge r_1~,
    \end{cases}
\end{equation}

\noindent where $r_2$ and $\gamma_1$ are the new free parameters because $r_3$ can be determined by imposing the continuity of $\xi(\Delta r)$. In this way, $r_3 = r_2(r_0/r_1)^{\gamma/\gamma_1}$. In Fig. \ref{tcf_civ} we present the resulting best fit for this model as the black line, corresponding to the best fit parameters: $r_1 = 654^{+100}_{-87}$ kpc, $r_2 = 4.70^{+1.60}_{-1.19}$ kpc, and $\gamma_1 = 1.78 \pm 0.10$. We again used the galaxy-galaxy autocorrelation measurements of $\gamma$ and $r_0$ \citep{Durkalec15}, and we considered the data points from \cite{Martin10}. When we follow the same reasoning from the arguments for the first break of $\xi(\Delta r)$, a second break implies that these smaller scales probe a different structure than the one we see at larger separations. This model indeed fits the data much closer than the simpler models that struggle at the smaller scales, which strongly supports the idea that we probe a different type of environment at the smaller scales probed by this survey ($\lesssim 10$ kpc).

\section{Discussion}

\begin{figure}
  \centering
  \includegraphics[width=1\columnwidth]{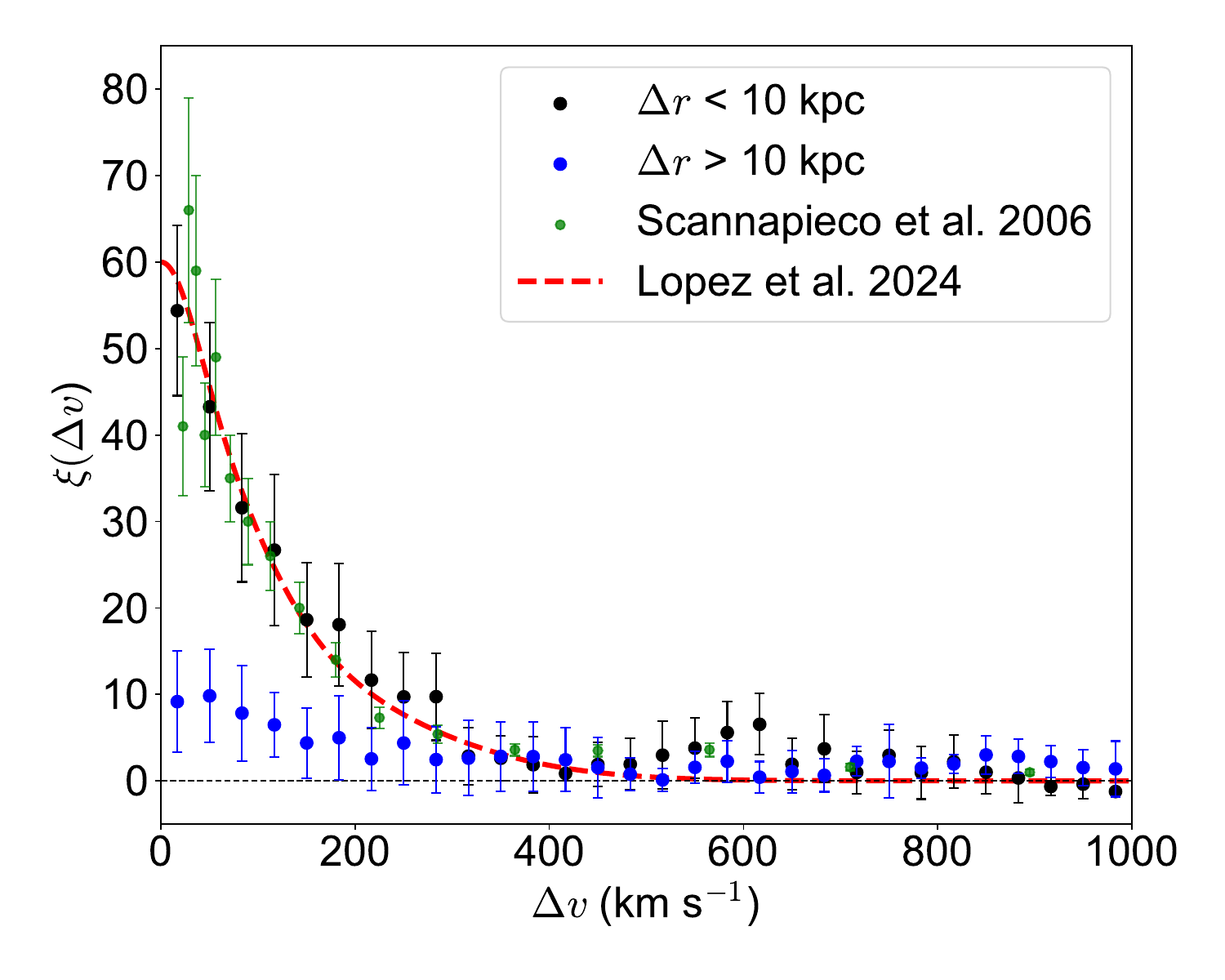}
  \caption{Two-point correlation function ($\xi(\Delta v)$) of \civ\ as a function of velocity across the two lines of sight for two different bins in separation, shown as the black and blue points for \sep\ $<10$ kpc and \sep\ $>10$ kpc, respectively. The green points correspond to the single line-of-sight measurements obtained by \citet{Scannapieco06}. The dashed red line shows a prediction from transverse correlations of \civ\ using gravitational arcs \citep{Lopez24}, adjusted to account for the different spectral and spatial resolution.}
\label{acf_v_civ}
\end{figure}

The connection between galaxies and \civ\ is well documented (e.g., \citealp{Adelberger05b, Martin10, Turner14, Bird16, Rudie19, Galbiati23, Banerjee23}). For this reason, our fits from Eq. \ref{eq:break1} and \ref{eq:break2} directly take the same correlation length from galaxies into account. Our correlation measurements at the higher separation bins lie exactly on top of the galaxy autocorrelations (Fig. \ref{tcf_civ}) that were also found by \cite{Martin10} when comparing with \cite{Adelberger05b}. This then suggests that \civ\ might be a good tracer of the way in which galaxies cluster at cosmic noon. We observe, however, that the strong relation with galaxies does not hold at $\Delta r \lesssim 500$ kpc, as seen in the flattening in of \xir\ in Fig. \ref{acf_r_civ} (this is the scale of the intermediate regime in Fig. \ref{tcf_civ}, where $\Xi(\Delta r)/\Delta r \propto 1/\Delta r$). It has been argued \citep{Scannapieco06, Martin10, Mintz22} that this flattening of $\xi$ provides constraints on the sizes of \civ\ enriched regions around galaxies. The argument is based on the fact that by decreasing \sep, we do not find a significant number of additional pairs of velocity components, which then implies that we probe distances smaller than these regions. In conclusion, the break that we find on large scales with the two models from Eq. \ref{eq:break1} and \ref{eq:break2} ($597 \pm 130$ kpc and $654^{+100}_{-87}$ kpc) agrees well with the result reported by \cite{Martin10} of  $r_1 = 600 \pm 214$ kpc.

On the other hand, at the smaller scales probed here, Fig \ref{acf_r_civ} shows that $\xi(\Delta r)$ increases again. We attribute this to the fact that at this scale, the autocorrelation is dominated by pairs of velocity components on the same absorption system. Therefore, this second break at $r_2 = 4.70^{+1.60}_{-1.19}$ kpc would provide limits on the sizes of individual \civ\ clouds within the CGM of galaxies, instead of whole absorption systems. Unfortunately, the sizes of these clouds are difficult to constrain. For instance, \cite{Rauch01} measured characteristic sizes of \ion{C}{iv} clouds in the range $r_0 = [3, 19]$ kpc, also using lensed quasars, but a different method to constrain sizes. On the other hand,~\citet{Lopez24} found rather smaller sizes ($\lesssim 1$ kpc) from kinematic arguments, but concentrated on the very strong \civ\ systems toward gravitational arcs. The existence of a break in $\xi$ at $r_2 = 4.70^{+1.60}_{-1.19}$ kpc, as shown in this work, strongly supports the idea of a coherence length of the cool-warm CGM at the scale of a few kiloparsec.

It is worth comparing our results with those for other species. Also using gravitational arcs, \cite{Afruni23} measured a coherence length of $4.2 ^ {+3.6} _ {-2.8}$ kpc by studying absorption systems of \ion{Mg}{ii} at lower redshifts ($z\approx1$). Interestingly, this result is well within $1\sigma$ from ours, even when we consider that \civ\ is more ionized and typically covers a larger cross section than \ion{Mg}{ii} (e.g., \citealp{Dutta21}). Independently, \cite{Rubin18} measured a lower length for the coherence length of $1.9$ kpc, which is still consistent with our result, but might also indicate a possible difference between the coherence length of these ions. Therefore, further work is needed to assess a possible difference between these coherence scales. 

It is also interesting to make a comparison with independent measurements of the clustering of \civ\ systems. To do this, we first projected \xvr\ in velocity. Fig. \ref{acf_v_civ} shows $\xi(\Delta v)$ in two bins around $\Delta r = 10$ kpc. Notably, the amplitudes of \xiv\ in the two bins are quite different, which is expected given the results of Section \ref{TCA}. This difference makes a direct comparison with single quasar line of sight results more difficult, however. At small separations, \xiv\ becomes comparable with the velocity clustering measured directly toward single quasar lines of sight (green points), obtained using data with a similar resolution at $z\approx2.2$~\citep{Scannapieco06}. Based on this similarity, we fit our \xiv\ points at $\Delta r < 10$ kpc using two Gaussians following \citet{Boksenberg15},

\begin{equation}
    \xi(\Delta v) = A_1 \exp \Bigg(-\dfrac{\Delta v^2}{2 \sigma_1^2}\Bigg) + A_2 \exp \Bigg(-\dfrac{\Delta v^2}{2 \sigma_2^2}\Bigg),
\end{equation}

\noindent where $A_{1,2}$ are the amplitudes, and $\sigma_{1,2}$ the corresponding standard deviations. We obtained a best fit with $A_1 = 26.2\pm15.3$, $\sigma_1 = 48.8\pm33$ \kms, $A_2 = 29.5\pm11.7$, and $\sigma_2 = 166.3\pm28.4$ \kms. These standard deviations are consistent with those obtained by \citet{Boksenberg15} at $\sigma_{1,2} = 35.5, 105$ \kms, suggesting that \xiv\ as measured by quasar pairs at small separations, where absorption profiles become more similar, is comparable to that measured using single lines of sight.

In an independent comparison, the dashed red line of Fig. \ref{acf_v_civ} shows the measurement obtained by \cite{Lopez24}. These authors computed a transverse \xiv\ for \ion{C}{iv} absorption systems using a different technique, namely integral field spectroscopy of giant gravitational arcs, over separations of $\Delta r \lesssim 10$ kpc. When we use the relation between $\sigma$ of $\xi$ for arcs and QSOs proposed by these authors, namely, $\sigma_{\text{arc}} = \sigma_{\text{QSO}} / \sqrt{N}$, where $N$ is the mean number of clouds per spatial resolution element in~\cite{Lopez24}, we obtain the relation shown by the dashed red line. This relation matches our measurements at $\Delta r < 10$ kpc remarkably well considering the different methods. We conclude that measurements of $\xi$ toward point-like and extended sources are indeed consistent with each other.

\section{Summary}
We assembled a sample of high-resolution spectra of 12 quasar pairs to study the clustering of \civ\ absorption systems around cosmic noon. Our results are listed below.
\begin{enumerate}
    \item We detected 141 \civ\ absorption systems at $z \approx 2$, in parallel quasar line of sights (A and B) spanning separations between 20 pc and up to 2.37 Mpc.
    \item We fit Voigt profiles to all velocity components, which resulted in a sample of 327 \civ\ velocity components on lines-of-sight A and 293 on lines-of-sight B, with a column density distribution that agreed with the known \civ\ statistics (Fig. \ref{fNz}).
    \item We computed the two-point correlation function of \civ\ velocity components ($\xi(\Delta v, \Delta r)$) using the \cite{Landy93} estimator (Fig. \ref{acf_r_civ}). We found that the amplitude of \xir\ flattens in the $\approx10-500$ kpc range, which confirms previous predictions of flattening for the \xir\ in this range \citep{Martin10, Mintz22}. 
    \item We computed the projected transverse correlation function ($\Xi(\Delta r)$) (Fig. \ref{tcf_civ}). We then tested models of broken power laws to estimate the coherence length of \civ\ absorption systems. For the larger scales we found a best fit at $r_1 = 654^{+100}_{-87}$ kpc, which we consider the scale of these metal-enriched regions.
    \item For the smaller scales ($\lesssim 10$ kpc), the correlation amplitude increases with decreasing separation. This suggests a different regime of the correlation function at scales $r_2 < 4.70^{+1.60}_{-1.19}$ kpc, which we interpret as a relevant scale for the coherence length of \ion{C}{iv} clouds.
    \item Our results offer the first comprehensive picture of the clustering of metal-enriched regions and shows that they are separately compatible with observations of \civ\ toward giant gravitational arcs (Fig. \ref{acf_v_civ}), previous measurements of cloud sizes for \ion{C}{iv}, and the coherence length of the CGM.
\end{enumerate}
While our results suggest a small coherence length of \ion{C}{iv} clouds, further work is necessary to measure more precise correlation amplitudes at separations $\lesssim 1$ kpc. To achieve this, more spectroscopic data on lensed quasars (or galaxies) are needed.  On larger scales, additional data on quasars pairs would significantly help us to constrain the scale of the metal-enriched regions, whose uncertainties are still on the order of hundreds of kiloparsecs.

\begin{acknowledgements}
We would like to thank the anonymous referee for their constructive feedback and comments. H. C., S. L. and N. T. acknowledge the support by FONDECYT grant 1231187. P. A. acknowledges support from ANID-Subdirección de Capital Humano/Doctorado Nacional/2022-21222110.
\end{acknowledgements}

\bibliographystyle{bibtex/aa}
\bibliography{bibtex/library.bib}

\begin{appendix}

\section{Integral Constraint}

A well known bias in correlation measurements that we did not address in the main text is the so-called "integral constraint". This comes from the fact that the redshift space of each quasar is limited. Then, when we measure positive correlation, this can cause a negative one somewhere else, but at the same time the real correlation can be assumed to be positive. As shown by \cite{Landy93}, the real correlation ($\xi^{real}$) is related to the measured one ($\xi$) by:

\begin{equation}
    1+\xi = \dfrac{1+\xi^{\text{real}}}{1+\xi_V},
\end{equation}

\noindent where $\xi_V$ is the integral constraint computed as follows:

\begin{equation}
    \xi_V = \int_V G(r)\xi^{\text{real}}(r) \text{d}^2V,
\end{equation}

\noindent with $G(r) = R_A R_B/n_{RR}$. As we do not know $\xi^{\text{real}}$ a priori, we only made a small correction following \cite{Tejos14},

\begin{equation}
\begin{split}
    \xi = (1+\tilde{\xi}_V)(1+\xi) - 1 \\\\
    \tilde{\xi}_V = \int_V G(r)\xi(r) \text{d}^2V,
\end{split}
\end{equation}

\noindent which is a small correction in this case ($\lesssim 1\%$), smaller than our uncertainties. We nonetheless applied this correction.

\section{Continuity of the projected transverse correlation function.} \label{cont_tcf}

Imposing the continuity of $\Xi$, at $\Delta r = r_2$, it is immediate that,

\begin{equation}
    \Xi (\Delta r) = 
    \begin{cases}
        A(r_3, \gamma)\Big(\dfrac{r_1}{r_2}\Big)^{\gamma/\gamma_1}\Delta r^{1-\gamma_1}, & \Delta r \le r_2 \\
        A(r_0, \gamma)r_1^{1-\gamma}, & r_2 < \Delta r < r_1 \\
        A(r_0, \gamma)\Delta r^{1-\gamma}, & \Delta r \ge r_1,
    \end{cases}
\end{equation}

\section{Additional fits to the projected transverse correlation function.}

\begin{figure}
  \centering
  \includegraphics[width=1\columnwidth]{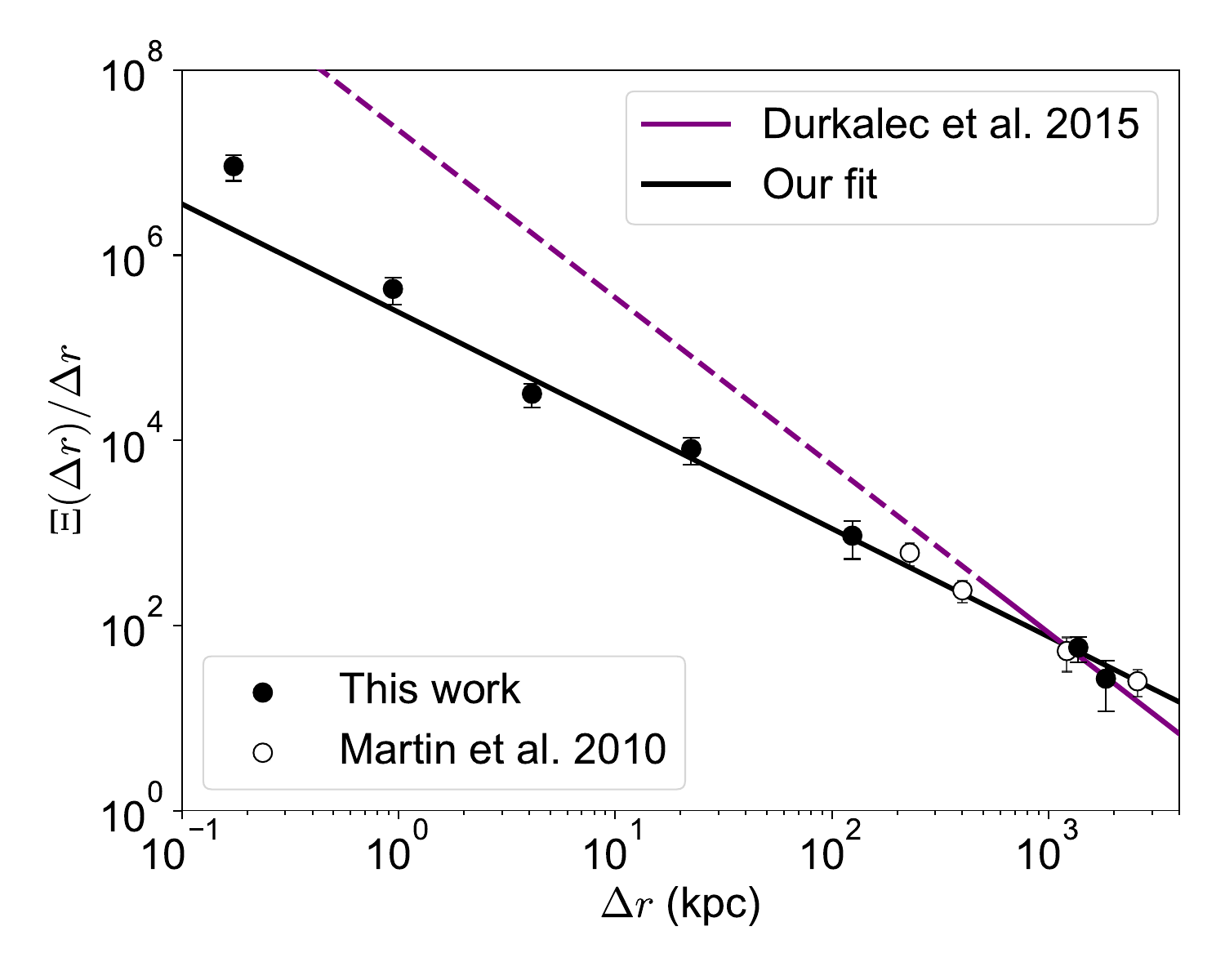}
  \caption{Projected transverse correlation function ($\Xi (\Delta r)$) for \ion{C}{iv}. Same as Fig. \ref{tcf_civ}, with the difference that here we fit a single power law of the form shown in Eq. \ref{eq:fit}. Best fit parameters: $r_0 = 4.45 \pm 0.56$ Mpc, and $\gamma = 1.17 \pm 0.05$.}
\label{tcf_civ1}
\end{figure}

\begin{figure}
  \centering
  \includegraphics[width=1\columnwidth]{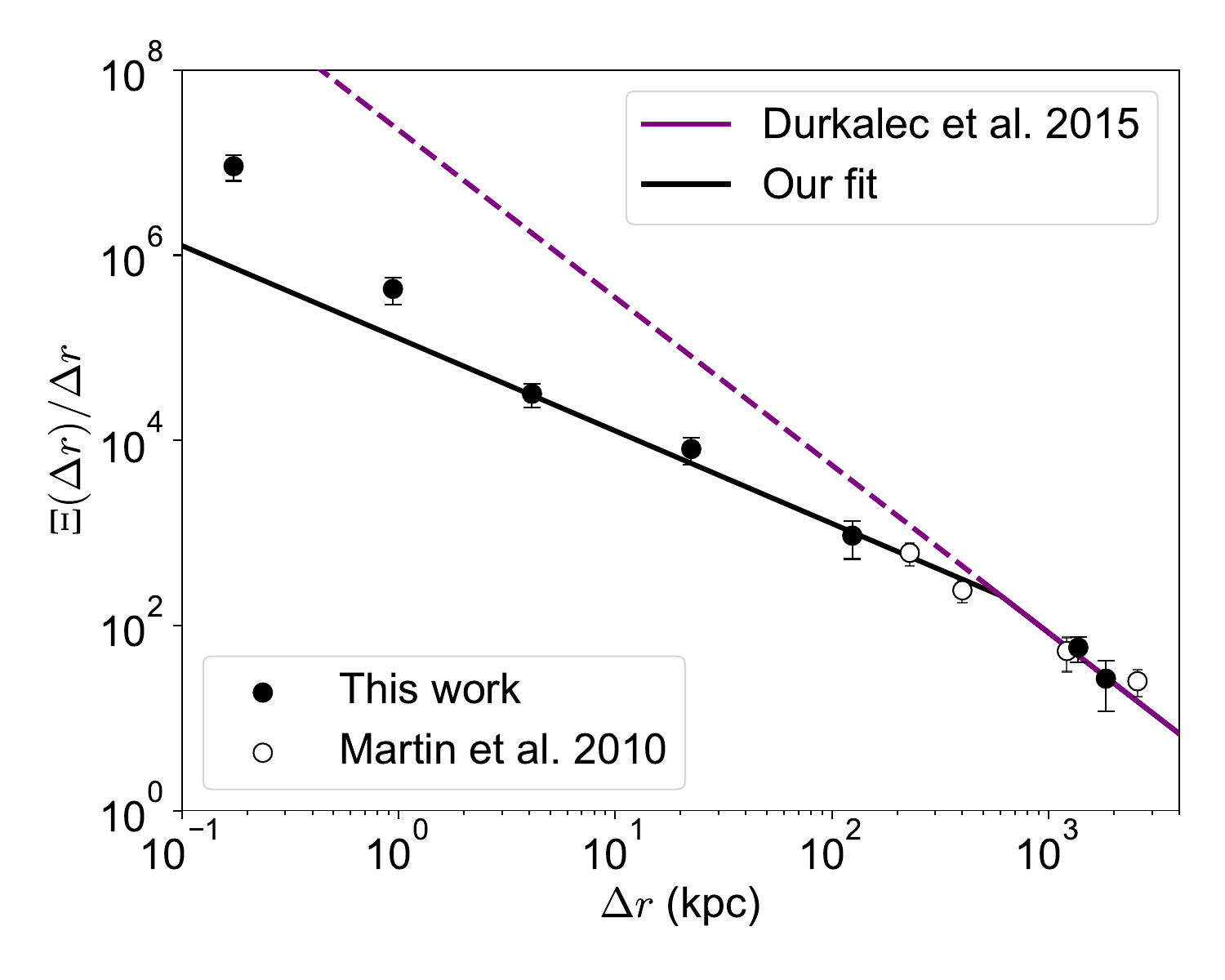}
  \caption{Projected transverse correlation function ($\Xi (\Delta r)$) for \ion{C}{iv}. Same as Fig. \ref{tcf_civ}, with the difference that here we fit a broken power law of the form shown in Eq. \ref{eq:break1}. Best fit parameter: $r_1 = 597 \pm 130$.}
\label{tcf_civ2}
\end{figure}

\end{appendix}

\end{document}